%% file: main.tex
\documentclass[runningheads,orivec]{llncs}
\usepackage[T1]{fontenc}
\usepackage{graphicx}
\usepackage[misc]{ifsym} %

\usepackage{subfig} %
\usepackage{booktabs}
\usepackage{bbding} %
\usepackage{tikz}
\usetikzlibrary{automata, backgrounds, calc, positioning, arrows.meta, decorations.pathmorphing,shapes,tikzmark,shapes.multipart}
\tikzset{
	->, %
	>=Stealth, %
	node distance=3cm, %
	every state/.style={thick, fill=gray!10}, %
	initial text=$ $, %
}
\usepackage[scaled]{beramono}
\input{macros}

\usepackage{listings}
\usepackage{wrapfig}
\usepackage{makecell}
\usepackage{multirow}

\usepackage{chngcntr}

\AtBeginDocument{
    \counterwithout{lstlisting}{section}
    
}

\begin{document}
\title{{\huge{\sweap}}: Reactive Synthesis for Infinite-State Integer Problems}
\titlerunning{\sweap: Reactive Synthesis for Infinite-State Integer Problems}

\author{
Shaun Azzopardi\inst{1}\textsuperscript{(\Letter)}\orcidID{0000-0002-2165-3698}\and
\\Luca {Di Stefano}\inst{2}\orcidID{0000-0003-1922-3151}%
\and
Nir Piterman\inst{3}\orcidID{0000-0002-8242-5357}%
}
\authorrunning{Azzopardi et al.}
\institute{
Dedaub, San Gwann, Malta ---
\email{shaun.a@dedaub.com}\and
\mbox{TU\,Wien,\,Institute\,of\,Computer\,Engineering,\,Treitlstraße\,3,\,1040\,Vienna,\,Austria}
\and
\!\!\mbox{University\,of\,Gothenburg\,and\,Chalmers\,University\,of\,Technology,\,Gothenburg,\,Sweden}
}
\maketitle              %
\begin{abstract}
Recent years have seen a significant increase in the interest in reactive synthesis from specifications that relate to infinite state spaces.
We present \sweap, a tool for synthesis of infinite-state Linear Integer Arithmetic reactive systems. 
\sweap implements a CEGAR approach, relying on state-of-the-art finite-state synthesis tools as black boxes to solve abstract synthesis problems.
\sweap supports most common input formalisms for infinite-state reactive-synthesis problems: Temporal Stream Logic Modulo Theories, Reactive Program Games, the bespoke input of the \issy tool, and our own bespoke input. 
We present a mature version of \sweap with novel features: a dual abstraction approach that improves its capabilities in proving unrealisability, support for nondeterministic and unbounded updates, more general initialization of variables, and equirealisable reductions for optimisation.
Experimental evaluation shows that \sweap outperforms its only competitor in this domain. 
\end{abstract}
\section{Introduction}

Reactive synthesis aims at generating correct-by-construction programs from high-level temporal specifications.
This problem is framed as a two-player game between a \emph{controller} and an adversarial \emph{environment}~\cite{DBLP:conf/popl/PnueliR89}. A program, then, is a winning strategy for the controller (if any exists). The traditional setting is that of synthesis over propositions. Given the success of state-of-the-art techniques in this finite domain, many in the community have started moving beyond, to infinite-state domains. Several tools and techniques have been developed, showing surprising effectiveness for an undecidable problem.

First attempts to tackle this problem relied on a counter-example guided abstraction refinement (CEGAR) approach --- a finite abstraction of the infinite-state problem is computed and fed to an LTL synthesis tool, and the strategy received is used to refine the abstraction or conclude un/realisability. Approaches that originally tackled problems defined in Temporal Stream Logic (\tsl)~\cite{10.1007/978-3-030-25540-4_35}, LTL modulo theories~\cite{MaderbacherBloem22}, and others~\cite{rodriguez25counter}, are of this kind.
By design, these focus on learning new safety properties of the underlying theory, relative to the predicates and updates in the objective.
Since the cost of finite LTL synthesis is high, these approaches start with essentially no abstraction and build it through refinement. In any case, the basic CEGAR setup proves insufficient for infinite-state problems where liveness is important: in this context, one may never learn enough properties to conclude that, e.g., incrementing an integer repeatedly will eventually make it greater than an arbitrary value. 

Later approaches targeting Reactive Program Games (\rpg) and \issyl improved on this by means of symbolic analysis of the game graph itself, relying on quantifier elimination to compute winning regions of the (infinite but symbolically represented) graph. 
They rely on \textit{acceleration} to avoid the infinite refinement that plagued  
CEGAR-based approaches~\cite{10.1145/3632899,DBLP:conf/cav/HeimD25}.

This paper concerns \sweap, a CEGAR-based tool for synthesising strategies for LTL games over infinite-state arenas. Our approach~\cite{Azzopardi_2025} goes beyond 
previous CEGAR-based work, by starting with a full predicate abstraction and automatically learning relevant liveness properties of the game.
The abstraction and the LTL objective are fed to an LTL synthesis tool in each CEGAR round. 
We rely on a novel binary encoding of predicates that reduces the size and computation time of predicate abstraction to polynomial rather than exponential. 
Thanks to this efficient encoding, the size of the LTL formula is no longer exponential in the number of predicates, reducing the cost of synthesis. 
This binary encoding enables \sweap to eagerly add relevant LTL formulas, to the abstraction, that describe the effect of state updates (inc/decrementing) on the predicate state.

In this paper, we present a mature version of \sweap supporting higher-level problem that include numeric inputs, uninitialised variables, and unbounded controller updates. Furthermore, we address \sweap's previous limitations with solving unrealisable problems: we describe how \sweap can dualise the problem by under-approximating control of the arena, rather than the standard over-approximating. We also describe how we implemented automatic translations of other languages into the \sweap, allowing better comparison between different approaches, along with several equirealisable reductions.
On benchmarks from literature, we show how these new features increase the capacity of \sweap and  that \sweap outperforms the main tool in this domain on literature benchmarks.

\section{The {\Large{\sweap}} Approach}

\paragraph{Problem formulation.}
\sweap solves infinite-state reactive synthesis problems described as LTL games. The game arena is a symbolic automaton with a finite number of control states. Transitions relate to input, output, and state variables. The objective is an LTL formula over control states and first-order predicates over all state variables, inputs, and outputs. 
Inputs and state variables may have infinite domains, while outputs are Boolean. 
State variables differ from inputs and outputs in that, unless updated explicitly (w.r.t. previous values), they preserve their value. 
Arena transitions describe conditions on the next value of state variables, with 
nondeterministic updates, including infinite-branching, resolved by the controller.

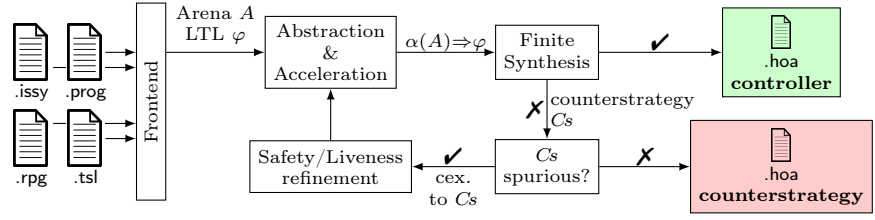
\begin{figure}[tb]
\centering
\begin{tikzpicture}[
auto,every node/.style={font=\scriptsize}
]

\node[draw, minimum height=8em] (front) {\rotatebox{90}{Frontend}};

\node[draw, right=4em of front,yshift=2em,align=center] (abs) {Abstraction\\\&\\Acceleration};
\node[draw, right=4em of abs, align=center] (syn) {Finite\\Synthesis};

\node[draw, fill=green!20, right=5em of syn,align=center] (real) {\reflectbox{\large\PaperPortrait}\\[-0.1em]\texttt{.hoa}\\\textbf{controller}};

\node at (front |- abs) [] (front-anchor) {} ;

\node[inner sep=1,fill=white,left=1.5em of front-anchor,label={[yshift=1.3pt]below:\texttt{.prog}}] (file-prog) {{\huge\reflectbox{\PaperPortrait}}} ;
\node[inner sep=1,fill=white,below=1em of file-prog,label={[yshift=3pt]below:\texttt{.tsl}}] (file-tsl) {{\huge\reflectbox{\PaperPortrait}}} ;
\node[inner sep=1,left=0.6em of file-prog,label={[yshift=3pt]below:\texttt{.issy}}] (file-issy) {{\huge\reflectbox{\PaperPortrait}}} ;
\node at (file-issy |- file-tsl) [inner sep=1,label={[yshift=1.1pt]below:\texttt{.rpg}}] (file-rpg) {{\huge\reflectbox{\PaperPortrait}}} ;

\node[draw, below=2em of syn, yshift=-0.4em, align=center] (conc) {$C\!s$\\ spurious?};
\node at (conc -| abs) [draw,align=center] (ref) {Safety/Liveness\\refinement};
\node at (conc -| real) [draw, fill=red!20,align=center] (unreal) {\reflectbox{\large\PaperPortrait}\\[-0.1em]\texttt{.hoa}\\\textbf{counterstrategy}};

\begin{scope}[->,>=latex]
    \draw (front.east |- abs) to [] node [above,align=center] {Arena $A$\\LTL $\varphi$} (abs) ;
    \draw (abs) to [] node [above,yshift=-3pt] {$\alpha({A})\mathord{\Rightarrow}\varphi$} (syn) ;
    \draw (syn) to [] node[left,xshift=2pt] {\XSolidBrush} node [right,xshift=-2pt,align=left] {counterstrategy\\$C\!s$} (conc) ;
    \draw (syn) to [] node[above,yshift=-3pt] {\CheckmarkBold} (real) ;

    \draw (conc) to [] node [above,yshift=-3pt] {\CheckmarkBold} node [below,align=center] {cex.\\to $C\!s$} (ref);
    \draw (conc) to [] node [above,yshift=-3pt] {\XSolidBrush} (unreal);

    \draw (ref) -- (abs);

    \draw (file-prog) -- (file-prog -| front.west) ;
    \draw (file-tsl) -- (file-tsl -| front.west) ;

\begin{scope}[on background layer]
\node [inner sep=0] at ($(file-rpg.east)+(0,0.2)$) (rpg-anchor) {};
\draw (rpg-anchor) -- (front.west |- rpg-anchor) ;

\node [inner sep=0] at ($(file-issy.east)+(0,-0.2)$) (issy-anchor) {};
\draw (issy-anchor) -- (front.west |- issy-anchor) ;
\end{scope}

\end{scope}
\end{tikzpicture}
\vspace*{-3pt}
\caption{Overview of \sweap's workflow.}\label{fig:workflow}
\vspace*{-10pt}
\end{figure}

\paragraph{CEGAR approach.}
\sweap takes a standard CEGAR approach~\cite{DBLP:journals/jacm/ClarkeGJLV03} (see Fig.~\ref{fig:workflow}). 
Given arena $A$ and objective $\varphi$,
\sweap creates the abstract arena $\alpha(A)$ via predicate abstraction~\cite{GS97cas}, and synthesises for $\alpha(A) \Rightarrow \varphi$
with a finite synthesis engine (Strix~\cite{DBLP:conf/cav/MeyerSL18} or SemML~\cite{DBLP:conf/tacas/KretinskyMPZ25,semml2}). 
By abstraction soundness, if a controller is found it is also valid for the concrete problem. 
Spurious counterstrategies (determined through invariant checking), are used to refine the abstraction for a new attempt.

\sweap innovates on the CEGAR approach for synthesis \cite{DBLP:conf/icalp/HenzingerJM03,MaderbacherBloem22} by using \emph{liveness} refinement, only resorting to safety refinement when the former is unsuccessful. This refinement relies on termination checking of the concrete behaviour induced by spurious lassos in abstract counterstrategies; and then refining the abstract problem with the learned terminating behaviour.

\begin{wraptable}{r}{0.44\linewidth}
\vspace{-1cm}
\begin{lstlisting}[style=sweapminimal,caption={\sweap example.},label={ex}]
INPUTS { i : natural }
OUTPUTS {}
STATE VARIABLES { x, y: natural }
CONTROL STATES { q: init }
OBJECTIVE { F y (*@$==$@*) 0 }
TRANSITIONS
 {q -> q [x (*@$>$@*) 0 & y (*@$>$@*) 0 
          # x(*@'@*) (*@$>=$@*) x - 1 &
             y(*@'@*) (*@$=$@*) y],
  q -> q [x (*@$=$@*) 0 & y (*@$>$@*) 0 
          # y(*@'@*) (*@$>=$@*) y - 1 &
             x(*@'@*) (*@$=$@*) i]   }
\end{lstlisting}
\vspace{-0.96cm}
\end{wraptable}
To manage the complexity of predicate abstraction (exponential) and synthesis (doubly exponential), \sweap uses a unique and natural binary encoding of predicates. Intuitively, LIA predicates are normalised into the form $t \leq c$, where $t$ is a term over variables, and $c$ a term over constant values. For each term,    predicates can be used to define mutually exclusive intervals that cover the number line. For example, given predicates $x \leq 0$, $x \leq 4$, and $x \leq 10$, we can define formulas over these predicates to state that $x \in (-\infty,0)$, $x \in [0,4)$, $x \in [4, 10)$, and $x \in [10,\infty)$. As these four statements are mutually exclusive, and their disjunction is a tautology, we can represent them using two Boolean variables.
By following these in the abstraction, rather that the single predicates, computing the abstraction becomes polynomial rather than exponential. Other CEGAR approaches create a fresh Boolean variable per predicate \cite{MaderbacherBloem22,rodriguez25counter,10.1007/978-3-030-25540-4_35}, making the Boolean LTL synthesis problem doubly exponential in the number of predicates. In our case, given the binary encoding, the complexity of the abstract synthesis problem is only exponential in the number of predicates. Importantly, \sweap further adds acceleration assumptions purely based on these intervals, encoding that incrementing a term infinitely often, without similar decrements, must lead to reaching higher intervals (and, similarly, decrementing leads to lower intervals)~\cite{Azzopardi_2025}. 

\begin{example}\label{ex:1}
Listing~\ref{ex} is an example of a \sweap problem. 
Transitions are guarded by two conditions, separated by \texttt{\#}, respectively over the current state (a guard) and the state change (an update). A transition is triggered when its guard holds, and consequently we require determinism over these guards. A primed variable $x'$ denotes the next value of variable $x$. The game has an LTL reachability goal stating that variable $y$ eventually has value $0$, to be achieved within the constraints of the arena. The arena has two transitions: the first is triggered when both $x$ and $y$ are greater than $0$, and lets the controller choose the next value of $x$ within the given constraints. Similarly, the second transition can only be taken when $x$ is $0$ and $y$ is greater than $0$, and allows some choice in the next value of $y$, but forces the next value of $x$ to match the current value of input variable $i$. Transitions describe how the state is allowed to evolve under certain conditions;
if no state change is allowed, e.g., when $x = y = 0$, the entire arena just stutters. \sweap determines this problem realisable, and returns a controller that simply chooses the smallest allowed value of $x$ and $y$ in each turn.
\end{example}

\paragraph{Other Features.} \sweap also supports one-shot finite synthesis over its input language, when all the types are finite (Boolean, or bounded intervals). Moreover, it provides arguments to verify returned controllers against the arena and objective through composing these in a nuXmv model. The user can also use this model to simulate the execution of the program against the strategy.

\section{Beyond Bounded Inputs and Updates (and more)}
The theory behind the prototype version of \sweap required fixed initial values of state variables, Boolean inputs, and a finite set of assignments for variable updates.
This setting can encode specifications equirealisable to any LIA specification in \tsl, \rpg, and \issyl (up to addition of extra variables), with excellent comparative performance and success rate~\cite{Azzopardi_2025}, but required manual translation of specification and the arena, which is error-prone \cite{fullpaper}.
We have since extended the syntax of \sweap's own input language to enable higher-level specifications (see Listing~\ref{ex}). 
Here we describe the novel features. Throughout we refer to predicates that refer only to state variables as \emph{state predicates} and those that also refer to inputs as \emph{input predicates}.

\paragraph{Uninitialized state variables.} 
With unintialised state variables, \sweap attempts to find a strategy that works for all possible initial values. When constructing the abstraction, \sweap determines all possible satisfiable combinations of state predicates. These options become an initial assumption in the abstraction formula, limiting the abstract environment to a legal initial choice of state predicates. 

\paragraph{Unbounded Inputs.} 
Numeric inputs are now a first-class feature and the predicate abstraction now also follows input predicates. 

Input predicates may relate inputs with state variables: in a na{\"i}ve approach, where the environment is unconstrained in its choice of input predicates, it would be allowed to choose illegal predicate combinations.
To maintain soundness, we construct the set of all legal input states as follows. First, we collect all satisfiable combinations of input predicates; then, for each combination, we perform quantifier elimination (QE), asking for a necessary condition on state variables for its satisfiability. We restrict the environment to this set of combinations and their conditions, enforcing local legality \cite{rodriguez25counter}. 

Given local legality, during spuriousness checking we can assume that inputs are always set in a way compatible with the input predicates chosen by the environment. Thus spuriousness is always due a wrong state predicate guess. However, such a wrong guess may be due to too coarse input predicates. Consider $x' = i$, where $x$ and $i$ are respectively state and input variables. The environment may wish $x = 7$ to be true in the next step, however it only knows $i \geq 0$. It sets this predicate in the current step and $x = 7$ in the next, as the transition is satisfiable. Spuriousness checking will find a mismatch, since not all models of $i \geq 0$ imply $x = 7$. Then we require refining of input predicates.

For safety refinement, given a counterexample, we previously relied solely on sequence interpolation~\cite{DBLP:conf/cav/McMillan06}. This constructs a sequence of formulas corresponding to the counterexample: the initial concrete counterexample state, formulas corresponding to the arena transitions and predicate choices in each time step, and a final formula representing the wrong guesses of the counterstrategy. Note that we do not include the concrete values of input variables in the counterexample, and ground all Boolean variables on their concrete values, giving a purely numeric formula. Variables appearing in these formulas are time-step indexed, giving a sequence of formulas $F_0,...,F_n, M$, where $M$ is the conjunction of the wrong predicate guesses in the last counterexample state. Interpolants are computed for each time-step $i$, as formulas over the common alphabets of $F_i$ and $F_{i+1}$. However, input variables are set freshly in each time-step and are never in the common alphabet: thus, interpolation can never discover input predicates.

When interpolation fails, we search for additional input predicates
that would allow the abstract environment to reach the predicate state ($M$) exposed by the counterexample. This resembles how QE is used in other work~\cite{rodriguez25counter}. We re-use the formulas constructed for interpolation: $f_{ce} = (F_0 \wedge ... \wedge F_n) \implies M$. As a quick first attempt, \sweap proceeds by performing QE over $\forall S_0,...,S_n. f_{ce}$, where $S_i$ is the set of state variables indexed by $i$. This gives us a formula over pure input predicates. 
If no fresh predicates are found, we search for predicates that relate inputs with state variables at each time-step, i.e. to relate the inputs and state variables at step $i$, we apply QE on the formula $\forall S_{\neq i}, I_{\neq i}. f_{ce}$, where $S_{\neq i}$ is the set of state variables not indexed by $i$, and similarly $I_{\neq i}$ for inputs. This gives us $n$ formulas. From each we extract a set of fresh input predicates, and choose the smallest non-empty such set to use in the next refinement step.

\paragraph{Unbounded Updates.}\label{par:updates} \sweap now supports underspecified updates to state variables, with the non-determinism resolved by the controller (see Listing~\ref{ex}). 
Canonical \sweap arenas are still as in the prototype: transitions with guards and a list of assignments for every state variable ($s := f(V)$ is an assignement, where $s$ is a state variable, and $f(V)$ a formula over all arena variables). The new high-level syntax is translated to this canonical form through a realisability-preserving transformation of the arena, that is essentially the same as we have previously described to encode numeric inputs in the arena state \cite{Azzopardi_2025}. 
For every transition that requires unbounded controller choice over variables, we redirect it to a fresh state that allows the controller to increment or decrement the variables, allowing it to continue to the original target only when the update condition holds. 
A modification of the LTL objective is also required, to ensure the evaluation of the formula stutters when the arena is in the introduced states, and a guarantee is added to ensure the controller eventually leaves such states.

\newcommand{\SweapStrix}{\texttt{sweap-strix}\xspace}
\newcommand{\SweapP}{\sweap^p\xspace}
\newcommand{\SweapD}{\sweap^d\xspace}

\newcommand{\PROGBENCHS}{95\xspace}
\newcommand{\TSLBENCHS}{65\xspace}
\newcommand{\RPGBENCHS}{86\xspace}
\newcommand{\ISSYBENCH}{115\xspace}
\newcommand{\NOTPROGBENCH}{266\xspace}

\section{Flipping Control of the Arena Abstraction}

By default the environment controls the state predicates. However, there are unrealisable problems which do not admit finite counterstrategies with this abstraction choice, leading to non-termination for our approach. We introduce a dual approach, that flips control of state variables to the controller, allowing for termination in these cases. We consider a simple illustrative example.

\vspace{-0.85cm}
{}\hfill
\begin{wraptable}{r}{0.45\linewidth}
\vspace{-0.5cm}
\begin{lstlisting}[style=sweapminimal,caption={Dualisation example.},label={ex-dual}]
... 
OBJECTIVE { (F q1) & (G x < 10) }
TRANSITIONS 
 { q0 -> q0 [!c1 & !c2 # x(*@'@*) = x + 1]
   q0 -> q0 [!c1 & c2 # x(*@'@*) = x - 1]
   q0 -> q0 [c1 & x < 0 # x(*@'@*) = x]
   q0 -> q1 [c1 & x >= 0 # x(*@'@*) = x]
   q1 -> q1 [true # x(*@'@*) = x + i] }
\end{lstlisting}
\vspace{-0.8cm}
\end{wraptable}
\begin{example}\label{ex:2}
Consider the problem in Listing~\ref{ex-dual},
where $c1$ and $c2$ are controller propositions, $i$ an integer input, and $x$ an integer state variable. In \texttt{q0} the controller can force any value of $x$, or move to \texttt{q1} (if $x$ is non-negative). At \texttt{q1} the environment chooses an $i$ to increment $x$ with.
The controller objective states that eventually the arena moves to \texttt{q1} and that $x$ is always bounded above by $10$.
\end{example}

This example is clearly unrealisable: if \texttt{q1} is reached, there is always an integer $i$ the environment can choose to make $x$ larger than $10$ (in fact, $i = 10$ always suffices). 
However, in our standard flow, where the environment must accurately emit the current predicate state in each turn, the winning counterstrategy is infinite. Consider $x = 0$ initially, and the controller decrements $x$, leading the environment to respond correctly with $x < 0$. At this point, if the controller chooses to increment $x$, the predicate abstraction offers two choices for the environment response: $x < 0$ or $x = 0$, i.e. precision is lost. In fact, responding precisely to any possible sequence of inc/decrements requires counting, and thus an infinite amount of state predicates.

Our solution is to provide a setting that switches control of state predicates to the controller. Essentially, in this dual approach, where the controller controls the state predicates (but the environment remains in control of input predicates) there is a finite counterstrategy: the environment can simply win by setting $i \geq 10$ when \texttt{q1} is reached, regardless of the controller choice for state predicates.

To support the dual setting, we change the interaction between the controller and the environment.
Normally, initially, the environment chooses initial values of state predicates. Then it chooses inputs followed by the controller choosing outputs. Then, in every step, the environment chooses predicates (based on the previously taken transition) followed by inputs, and the controller replies with outputs.
Conceptually, a dual interaction instead includes a choice by the controller for values of state predicates, a choice by the environment of inputs, and another choice by the controller for outputs and predicates (repeatedly in every step). 
We support this by moving the choice of state predicates to the previous time step.
Thus, initially, the controller chooses values for state predicates. 
Then, in every step, the environment chooses inputs (including input predicates), the controller chooses outputs and updates the state predicates based on the transition. 
It follows that arena transitions must be executed across time steps.

To enable this in the abstract problem, we change the produced LTL formulas by adding one next operator ($X$) before any occurrence of input or output variables, and input predicates. 
Namely, if the original formula is $\varphi$ and the description of the arena is $\alpha(A)$, %
the new formula for finite synthesis becomes $\alpha^{dual}(A) \wedge \varphi^{dual}$, with the modifications described. 
Notice that, in this new problem, it is the responsibility of the controller to obey the constraints of the arena.
Furthermore, the initial choices of inputs and outputs (at time 0) do not play a role in the evaluation of such formulas.
In the case of undetermined initial values, we include an additional interaction between the environment and the controller at time 0.
We check which initial predicate states are feasible and add enough Boolean inputs to the environment so that it can force one such state.
We then allow the controller, as explained, to choose the actual predicates.
We enforce that the controller copies correctly these values by adding a conjunct $\varphi_{init}$. 
Finally, input predicates, which are determined only by the environment, are also moved to the correct time step (i.e.,~scoped by $X$).

We force the environment to choose only legal combinations of input predicates, after the controller acts, by the formula $G(\varphi_{inputs})$, where $\varphi_{inputs}$ is a disjunction over all legal input predicate combinations, and obtain the final specification $G(\varphi_{inputs}) \implies \varphi_{init} \wedge \alpha^{dual}(A) \wedge \varphi^{dual}$. 

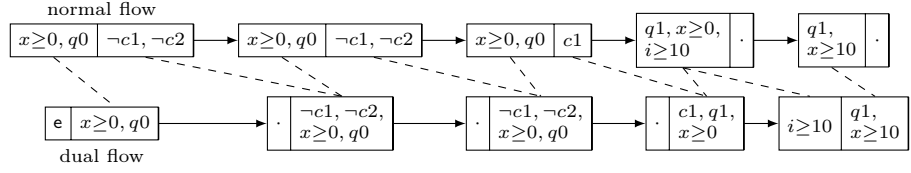
\begin{figure}[t]
\centering
\begin{tikzpicture}[
auto,every node/.style={font=\scriptsize}
]

\node [label=above:{normal flow},rectangle split,rectangle split horizontal,rectangle split ignore empty parts,draw] (normal1) {$x\mathord{\geq}0,q0$\nodepart{two}$\neg c1,\neg c2$};
\node [right=1.8em of normal1, rectangle split,rectangle split horizontal,rectangle split ignore empty parts,draw] (normal2) {$x\mathord{\geq}0,q0$\nodepart{two}$\neg c1,\neg c2$};
\node [right=1.8em of normal2, rectangle split,rectangle split horizontal,rectangle split ignore empty parts,draw] (normal3) {$x\mathord{\geq}0,q0$\nodepart{two}$c1$};
\node [right=1.8em of normal3, rectangle split,rectangle split horizontal,rectangle split ignore empty parts,draw,align=left] (normal4) {$q1,x\mathord{\geq}0,$\\$i\mathord{\geq}10$\nodepart{two}$\cdot$};
\node [right=1.8em of normal4, rectangle split,rectangle split horizontal,rectangle split ignore empty parts,draw,align=left] (normal5) {$q1,$\\$x\mathord{\geq}10$\nodepart{two}$\cdot$};

\node [label=below:{dual flow},below=2em of normal1, rectangle split,rectangle split horizontal,rectangle split ignore empty parts,draw] (dual1) {$\texttt{e}$\nodepart{two}$x\mathord{\geq}0,q0$};
\node at (dual1 -| normal2) [rectangle split,rectangle split horizontal,rectangle split ignore empty parts,draw,align=left
] (dual2) {$\cdot$\nodepart{two}$\neg c1,\neg c2,$\\$x\mathord{\geq}0,q0$};
\node at (dual1 -| normal3) [rectangle split,rectangle split horizontal,rectangle split ignore empty parts,draw,align=left] (dual3) {$\cdot$\nodepart{two}$\neg c1,\neg c2,$\\$x\mathord{\geq}0,q0$};
\node at (dual1 -| normal4) [rectangle split,rectangle split horizontal,rectangle split ignore empty parts,draw,align=left] (dual4) {$\cdot$\nodepart{two}$c1,q1,$\\$x\mathord{\geq}0$};
\node at (dual1 -| normal5) [rectangle split,rectangle split horizontal,rectangle split ignore empty parts,draw,align=left] (dual5) {$$\\$i\mathord{\geq}10$\nodepart{two}$q1,$\\$x\mathord{\geq}10$};

\begin{scope}[->,>=latex]
    \draw (normal1) -- (normal2);
    \draw (normal2) -- (normal3);
    \draw (normal3) -- (normal4);
    \draw (normal4) -- (normal5);
    \draw (dual1) -- (dual2);
    \draw (dual2) -- (dual3);
    \draw (dual3) -- (dual4);
    \draw (dual4) -- (dual5);
\end{scope}
\begin{scope}[dashed,-]
    \draw (normal1.one south) -- (dual1.two north);
    \draw (normal1.two south) -- (dual2.two north);
    \draw (normal2.one south) -- (dual2.two north);
    \draw (normal2.two south) -- (dual3.two north);
    \draw (normal3.one south) -- (dual3.two north);
    \draw (normal3.two south) -- (dual4.two north);
    \draw (normal4.one south) -- (dual4.two north);
    \draw (normal4.one south) -- (dual5.one north);
    \draw (normal5.one south) -- (dual5.two north);    
\end{scope}
\end{tikzpicture}
\caption{A trace from the abstraction of Listing~\ref{ex-dual} and its dualised counterpart.}\label{fig:dual-trace}
\end{figure}

Figure~\ref{fig:dual-trace} shows how an abstract trace for Example~\ref{ex:2} changes between the normal and the dual flow, when following the predicates: $x \geq 0$, $x \geq 10$, and $i \geq 10$. For simplicity, we write $x \geq 0$ for $x \geq 0 \land \neg(x \geq 10)$, and ignore $i \geq 10$ until \texttt{q1} is reached.
Each trace step contains two boxes 
for the environment and controller's choices in each timestep, respectively. When certain choices do not affect the objective or abstraction, we simply write $\cdot$ in their place. For the dual flow, we introduce fresh propositions to let the environment choose the initial arena state: here we introduce the environment proposition $e$ which when set forces the controller to choose $x\geq0$ and \texttt{q0}.
Note that state predicates (e.g., $x \geq 0$) stay at the same time step, but in the dual flow are chosen by controller. Controller outputs ($c1, c2$) and original environment inputs
($i$) are instead pushed to the next state. Hence, if the LTL objective had an assumption $G(q1 \rightarrow (i \leq  0))$, then for the dual problem this would be transformed into $G(q1 \rightarrow X(i\geq 0))$, to maintain the semantics and preserve (un)realisablity.

\section{Equirealisable Translations and Reductions}

\sweap now handles other input languages used by the community, specifically Temporal Stream Logic (\tsl), Reactive Program Games ({\rpg}s), and the \issy input format. 
Problems in each format can be translated to equirealisable problems in the other formats, but they differ in style. 
In this section, we detail how \sweap transforms these into its format, while maintaining equirealisability. 
We also present equirealisable reductions \sweap performs to reduce game size.

\paragraph{Reactive Program Games (\rpg).}
A problem described in \rpg is a deterministic symbolic arena with an explicit objective over arena control states (safety, reachability, B\"uchi, or parity). It includes numeric inputs with a finite set of assignments. The translation of the arena defined to \sweap's arena format is thus simply a parsing and compilation problem. The \rpg objectives are transformed into equivalent LTL objectives over arena states in the expected manner.

\paragraph{Temporal Stream Logic (\tsl).}
\tsl problems are LTL formulas, without any arena, over predicates and a finite set of assignments. 
To support assignments, we create an arena that allows the controller to choose among them. To keep the predicate abstraction small, rather than considering all the possible assignment combinations, we partition assignments based on a cone-of-influence analysis. Assignments depending on inputs are placed in the first partition. 
Each arena control state corresponds to a partition, and transitions from it to choosing a combination of assignments in the partition. 
The controller's non-deterministic choice is resolved with minimal fresh outputs. The \tsl formula is modified to serve as the objective, by replacing explicit references to each assignment $u = \texttt{v := expr}$, with where $\texttt{v}$ is updated at state $s$ by a statement $\neg s \mathbin{U} (s \wedge f_u)$, where $f_u$ is a formula over the fresh outputs that trigger $u$ in the state $s$ ($s$ corresponds to the partition $u$ is in). Predicates are handled similarly, so that they are only evaluated in the initial state of the arena, preserving equi-realisability.

\paragraph{\issy input format.}
An \issyl problem is a list of formula objectives and games, with transitions guarded by update formulas. Formula objectives cannot talk about game arena states, but range over input and state variables, and also over primed state variables, representing their next-state value.
These formulas and games each express their own objectives, and are interpreted conjunctively. 

We apply some optimisations: (i) Boolean state variables appearing only primed are transformed into (unprimed) controller propositions; and (ii) numeric state variables that are never related with other variables have their predicates given a binary encoding, and the fresh controller propositions are used instead.

\issyl problems do not require an explicit game arena. We check formula-only problems with Spot~\cite{DBLP:journals/corr/abs-2206-11366}, to identify whether an arena can be extracted directly from an objective, or whether an arena, such as that described for TSL, can be constructed. When these are not possible, we construct a dummy arena that allows the controller to choose any next value of any state variable.

When arenas are explicitly defined, we transform these into one \sweap arena. 
For each arena, we normalise transition conditions into DNF form. Unsatisfiable disjuncts are removed, while the update conditions and guards are separated.
Then, for each state, we get a set of guard and update-condition pairs.
By \issy semantics, the disjunction of the guards here must be a tautology. Given our optimisations (see (i) above), this may no longer be the case: we resolve this by a transition to a fresh losing state for the uncovered part of the domain. 
In the presence of multiple games, we compute their cross-product. Then, we resolve any non-determinism in the transition guards with fresh controller propositions, at which point we get a \sweap arena. Game arena objectives are expressed equivalently in LTL and conjuncted to the formula objectives. Objectives need to be massaged, replacing any predicate that references primed variables into their next-state representation (e.g., $x > x'$ becomes $X(x_\mathit{prev} > x)$).

\paragraph{Equirealisable Reductions.} \sweap performs several reductions to simplify games into equirealisabile ones, resulting in smaller synthesis problems.

When there is a non-deterministic choice of next variable value, \sweap analyses whether the choice is relevant to the game. It is relevant if the objective predicates over the variable, or if a data-flow analysis from the relevant state determines that it is accessed before being assigned a fresh value. Irrelevant variables from a state are stuttered, or, initially, given a default value.

\sweap also employs a reachability analysis, removing any unreachable arena control states. Furthermore, it computes the arena's strongly-connected components (SCCs), using these to optimise acceleration formulas. SCCs are also used during the translation of \issyl and \rpg problems to detect further losing states (e.g., states from which a B{\"u}chi objective is not reachable). Once a losing state is reached, the game stops and the arena state stutters.

\section{Experimental Evaluation}

In this section, we evaluate the presented version of \sweap\footnote{The development repository for \sweap is: \url{https://github.com/shaunazzopardi/sweap}.}, and how it compares to the state-of-the-art tool \issy~\cite{DBLP:conf/cav/HeimD25}. \issy is the only other available infinite-state synthesis tool that handles problems unsolvable by safety refinements alone, and supports the same formalisms as \sweap. 

\sweap refers to a virtual portfolio approach: we attempt to solve the same problem as-is and dualised, always using SemML as the finite synthesis engine, and then take the fastest time among the two. When both configurations reach a verdict, these are consistent across all experiments.
All experiments described in this section were run on a GNU/Linux computing cluster equipped with AMD EPYC 9355 processors.
Each task was given 4 CPU cores, a time limit of 10 minutes, and a memory limit of 64 GiB. We run \issy by using a container image obtained from its official Github repository.\footnote{\url{https://github.com/phheim/issy/}, revision \texttt{423bd1a}, as instructed by its developers.}

\begin{table}[t]
\scriptsize
\caption{Results of comparative evaluations of \sweap against \SweapStrix and \issy.
Columns report the number of \textbf{S}olved problems, \textbf{F}astest, and \textbf{U}nique solutions.}\label{tbl:compare}

\hfill
\subfloat[All problems in \sweap's input format.\label{fig:compare-sweap}]{
\begin{tabular}{l @{\hspace{0.8em}} c @{\hspace{0.8em}}c @{\hspace{0.8em}}c}
\makecell{Tool} & \makecell{S} & \makecell{F} & \makecell{U} \\\toprule
\SweapStrix & 65 & 41 & 0 \\
\sweap & \bf{84} & \bf{42} & \bf{19}\\
\bottomrule
\\
\multicolumn{4}{c}{Breakdown of 84 solved by \sweap:}\\
\toprule
\multicolumn{3}{l}{only {\bf base}:} & 52\\
\multicolumn{3}{l}{only {\bf dual}:} & 12\\
\multicolumn{3}{l}{{\bf base} faster than {\bf dual}:} & 8 \\
\multicolumn{3}{l}{{\bf dual} faster than {\bf base}:} & 12\\
\bottomrule
\end{tabular}
}
\hfill
\subfloat[Realisable problems in \issyl, \tsl, and \rpg.\label{fig:compare-real}]{
\begin{tabular}{l l @{\hspace{0.8em}} c @{\hspace{0.8em}}c @{\hspace{0.8em}}c}
\makecell{Format} & \makecell{Tool} & \makecell{S} & \makecell{F} & \makecell{U} \\\toprule
\multirow{2}{*}{\issyl}
& \issy  & 42 & 18 & 2 \\
& \sweap & \textbf{70} & \textbf{53} & \textbf{40} \\\midrule
\multirow{2}{*}{\rpg}
& \issy  & 52 & \textbf{34} & 2 \\
& \sweap & \textbf{57} & 25 & \textbf{7} \\\midrule
\multirow{2}{*}{\tsl}
& \issy  & 32 & 17 & 2 \\
& \sweap & \textbf{38} & \textbf{21} & \textbf{8} \\
\bottomrule
\end{tabular}}
\hfill
\subfloat[Unrealisable problems in \issyl, \tsl, and \rpg.\label{fig:compare-unreal}]{
\begin{tabular}{l@{\hspace{1em}} l @{\hspace{1em}} c @{\hspace{1em}}c @{\hspace{1em}}c}
\makecell{Format} & \makecell{Tool} & \makecell{S} & \makecell{F} & \makecell{U} \\\toprule
\multirow{2}{*}{\issyl}
& \issy     & 20 & \textbf{14} & 3 \\
& \sweap    & \textbf{21} & 5 & \textbf{4} \\\midrule
\multirow{2}{*}{\rpg}
& \issy  & 16 & \textbf{14} & 0 \\
& \sweap & \textbf{20} & 6 & \textbf{4} \\\midrule
\multirow{2}{*}{\tsl}
& \issy  & 14 & \textbf{10} & 0 \\
& \sweap & \textbf{17} & 7 & \textbf{3} \\
\bottomrule
\end{tabular}}
\hfill
\vspace*{-20pt}
\end{table}

\smallskip\noindent\emph{Evaluation of novel features.}
First, we consider the new version of \sweap alongside a configuration 
\SweapStrix
that uses Strix as the finite synthesis engine and does not use dualisation. This configuration roughly corresponds to the capabilities that the tool had in its prototype stage, but includes optimisations mentioned in this paper.
We perform our comparison over a collection of \PROGBENCHS benchmarks written in \sweap's input language \cite{Azzopardi_2025}.
72 of these are realisable. 
The results are summarised in Table~\ref{fig:compare-sweap}.
The new version of \sweap solves 84 problems, 19 more than \SweapStrix, and is faster on 42 of them.
\SweapStrix manages to be faster on 41 problems, but there are no problems that it can solve and \sweap cannot.
We note that the prototype version of \sweap required at least 20 minutes to solve a similar amount of these benchmarks as \SweapStrix~\cite{Azzopardi_2025}.
Specifically, we are able to solve 7 more problems than \SweapStrix just by using SemML;
the other 12 extra problems are solved through dualisation, which is also faster on 8 of the 20 problems solved by both workflows. Note that this represents 52\% of all 23 unrealisable benchmarks in this set.

\begin{figure}[pt]
    \centering
    \subfloat[\sweap vs \SweapStrix.\label{fig:strix}]{\includegraphics[width=0.49\linewidth,alt={Scatter plot comparing performance of the new version of Sweap (in a virtual portfolio configuration) with its prototype, based on Strix and without dualisation.}]{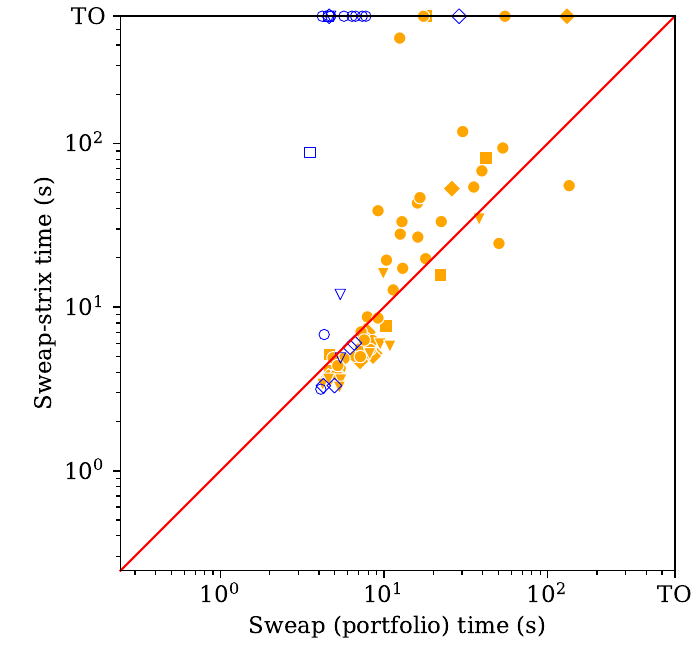}}
    \subfloat[\sweap vs \issy, \issyl benchmarks.\label{fig:issy-issy}]{\includegraphics[width=0.49\linewidth,alt={Scatter plot comparing performance of the new version of Sweap (in a virtual portfolio configuration) with Issy, on Issy problems.}]{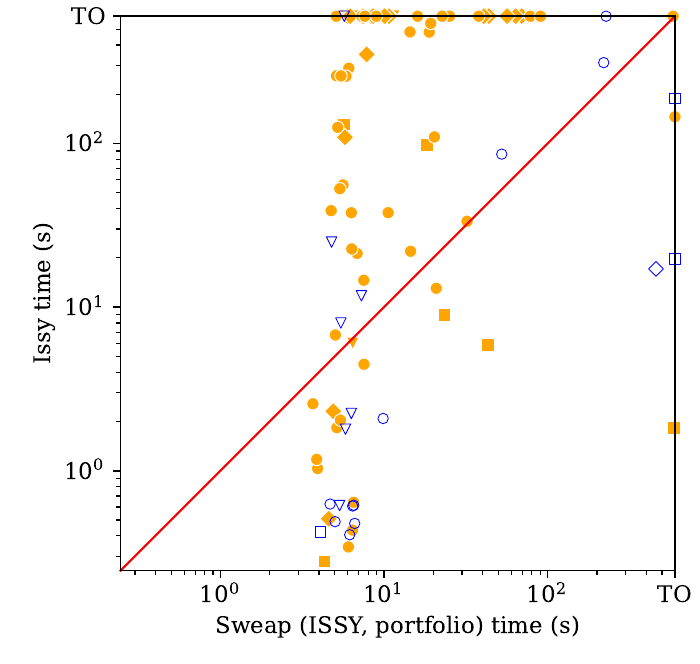}}

    \subfloat[\sweap vs \issy, \rpg benchmarks.\label{fig:issy-rpg}]{\includegraphics[width=0.49\linewidth,alt={Scatter plot comparing performance of the new version of Sweap (in a virtual portfolio configuration) with Issy, on RPG problems.}]{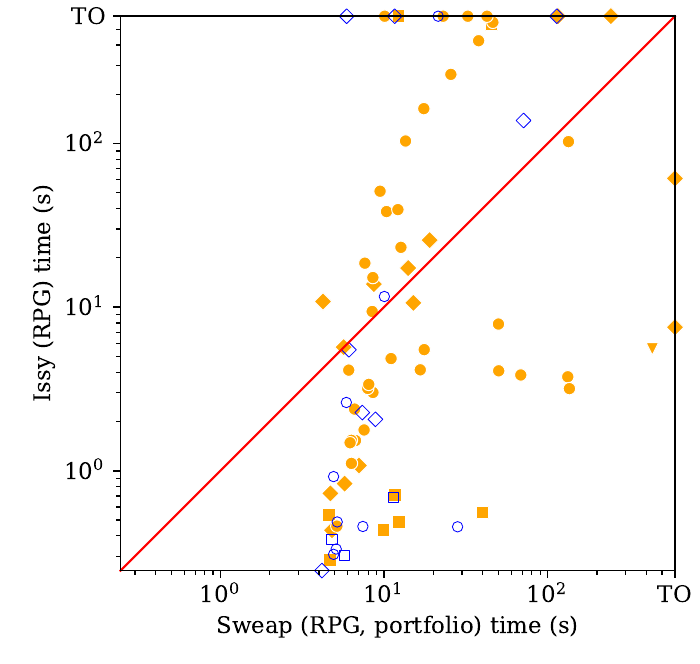}}
    \subfloat[\sweap vs \issy, \tsl benchmarks.\label{fig:issy-tsl}]{\includegraphics[width=0.49\linewidth, alt=alt={Scatter plot comparing performance of the new version of Sweap (in a virtual portfolio configuration) with Issy, on TSLMT problems.}]{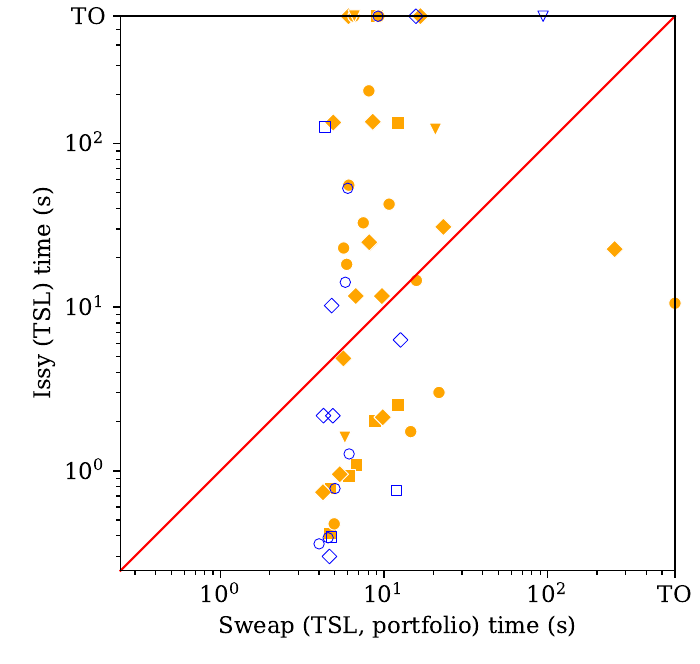}}
    
    \includegraphics[width=0.80\linewidth,alt={Legend for above scatter plots.}]{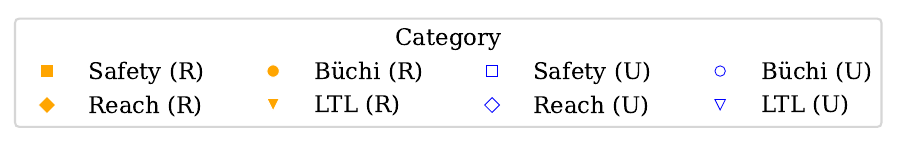}
    \caption{Scatter plots. Each marker corresponds to a synthesis task that both tools carried out without runtime errors.
    Execution times for our new tool are always mapped to the hoizontal axis.
    Filled-out and outlined markers correspond to \textbf{R}ealisable and \textbf{U}nrealisable problems, respectively.
    The timeout TO was set to 10 minutes. Details are provided in Appendix~\ref{apx:results}.}\label{fig:scatter}
\end{figure}

The scatter plot in Fig.~\ref{fig:strix}
compares the runtimes on experiments between \SweapStrix and \sweap. It shows that \sweap's superiority over \SweapStrix comes from solving unrealisable problems, which are hard without the novel dualisation feature, and from finding faster solutions to mildly complex problems, most of them with Büchi objectives. \SweapStrix remains faster on small problems, where the overhead of SemML may be overkill, possibly a sign that \sweap may benefit from future optimisations to reduce startup times.

\paragraph{Comparative evaluation with \issy.}
We collect a total of \NOTPROGBENCH reactive synthesis problems
over Linear Integer Arithmetic from the literature~\cite{10.1007/978-3-662-49674-9_12,10.1145/3632899,Azzopardi_2025,DBLP:conf/cav/HeimD25,DBLP:conf/cav/SchmuckHDN24,heim2026}
and try to synthesise for them using \sweap and \issy. We wish to compare how they perform on the \emph{synthesis problem}, rather than simply returning a verdict without a strategy. However, \issy does not produce counterstrategies, and thus we can only compare on realisable problems. For completeness, we also compare performance on unrealisable problems separately, but we stress that in this case \issy is solving a simpler problem, by only returning a verdict.

Out of \NOTPROGBENCH problems, \ISSYBENCH are expressed in \issyl (89 realisable), \RPGBENCHS in \rpg (66 realisable), and
\TSLBENCHS in \tsl (45 realisable).
Tables~\ref{fig:compare-real}--\ref{fig:compare-unreal} summarise the results for the two experiments.
For realisability, \sweap is able to solve more problems than \issy across all formats and find more unique solutions. \issy performs comparatively well, and is faster on more realisable RPG problems. Dualisation does not contribute significantly on realisable problems. For unrealisability, \sweap performs relatively well, even though it is solving a harder problem. Dualisation is helpful here, uniquely solving 5 \issyl, 8 \rpg, and 11 \tsl problems, and is  faster on average than the default configuration.
Note that whenever both tools manage to solve a problem, their verdicts are always consistent.

Fig.~\ref{fig:issy-issy}--\ref{fig:issy-tsl} compare execution times for the two tools. 
Though describing conceptually different comparisons, we show the results for both realisable and unrealisable problems together, due to space constraints. In general, \sweap incurs a certain overhead at startup, always requiring at least a few seconds to reach a verdict, but does better on larger problems.

We note that the \issy approach is less memory intensive than ours, due to our reliance on finite-synthesis tools. This is shown in Heim and Dimitrova's evaluation of \issy against the prototype version of \sweap \cite{DBLP:conf/cav/HeimD25,heim2026}, where only 4~GiB/6~GiB of memory were allocated. \sweap's default reliance on SemML further increases \sweap memory requirements to at least 40~GiB of memory.

\section{Conclusion}

We presented a mature version of our reactive synthesis tool \sweap. The tool supports all the major input formats
that recently emerged from infinite-state synthesis research, and its bespoke format has been extended to support
high-level specifications.
With its novel features and optimisations, the new version of \sweap significantly outperforms its earlier prototype~\cite{Azzopardi_2025}
and compares positively against the state-of-the-art tool \issy~\cite{DBLP:conf/cav/HeimD25}.

We believe we can optimise our CEGAR loop by manipulating the abstract game structures directly, instead of formulating them through fresh LTL formulas at every iteration. 
In~\cite{Azzopardi_2025} we argued that our technique may generalise to linear real arithmetic (LRA), and we are investigating a practical implementation.

\begin{credits}

\subsubsection{\ackname}
LDS is supported by the European Union (under grant agreement ID 101212818) and NP is supported by Swedish research council (VR) project (No. 2020-04963) and the Wallenberg AI, Autonomous Systems and Software Program (WASP) funded by the Knut and Alice Wallenberg Foundation.
\end{credits}

\clearpage
\bibliographystyle{splncs04}
\bibliography{references}

\clearpage
\appendix

\section{Experimental Results}\label{apx:results}
Table~\ref{tab:breakdown} reports a detailed breakdown of \sweap's results on \issyl, \rpg, and \tsl benchmarks, which we summarised in Tables~\ref{fig:compare-real}--\ref{fig:compare-unreal}.
Table~\ref{tab:results-prog} 
and
Tables~\ref{tab:results-issy}--\ref{tab:results-tsl} describe the result of each experiment in our comparative evaluation against
\SweapStrix and \issy, respectively.
The comparison against \issy is split in three tables, according to the benchmarks' input languages (\issyl, \rpg, or \tsl). 
Each row in each table reports the name of a benchmarks, its realisability ($\bullet$ marks realisable benchmarks), and synthesis times for
\issy and \sweap. An empty cell denotes timeout, which was set at 10 minutes. OOM denotes that a tool has exceeded its memory limit (64 GiB).
ERR denotes that the tool encountered a runtime error. In \issy experiment, we invoked the tool with the flag \verb|--synt| (to perform synthesis rather than just realisability).

\begin{table}[!b]
\centering\footnotesize
\caption{Breakdown of \sweap results on \issyl, \rpg, and \tsl benchmarks.}\label{tab:breakdown}
\begin{tabular}{cclr}
\toprule
\makecell{Format} & \makecell{Solved by\\\sweap} & \multicolumn{2}{c}{Breakdown}\\
\midrule
\multirow{4}{*}{\issyl} & \multirow{4}{*}{91}
  & {only {\bf base}:} & 61 \\
& & {only {\bf dual}:} & 5 \\
& & {{\bf base} faster than {\bf dual}:} & 8 \\
& & {{\bf dual} faster than {\bf base}:} & 17 \\
\midrule
\multirow{4}{*}{\rpg} & \multirow{4}{*}{77}
  & {only {\bf base}:} & 51 \\
& & {only {\bf dual}:} & 8 \\
& & {{\bf base} faster than {\bf dual}:} & 4 \\
& & {{\bf dual} faster than {\bf base}:} & 14 \\
\midrule
\multirow{4}{*}{\tsl} & \multirow{4}{*}{55}
  & {only {\bf base}:} & 36\\
& & {only {\bf dual}:} & 11\\
& & {{\bf base} faster than {\bf dual}:} & 2\\
& & {{\bf dual} faster than {\bf base}:} & 6\\
\bottomrule
\end{tabular}
\end{table}

\begin{table}[t]
\caption{Results on Sweap benchmarks (\SweapStrix vs \sweap).}\label{tab:results-prog}
\centering\tiny
\input{tables/table_sweap-pf_sweap-strix}
\end{table}

\begin{table}[t]
\caption{Results on \issyl benchmarks.}\label{tab:results-issy}
\centering\tiny
\input{tables/table_sweap-issy-pf_issy3}
\end{table}

\begin{table}[t]
\caption{Results on \rpg benchmarks.}\label{tab:results-rpg}
\centering\tiny
\input{tables/table_sweap-rpg-pf_issy3-rpg}
\end{table}

\begin{table}[t]
\caption{Results on \tsl benchmarks.}\label{tab:results-tsl}
\centering\tiny
\input{tables/table_sweap-tsl-pf_issy3-tsl}
\end{table}

\end{document}

%% file: macros.tex
\usepackage{wrapfig}
\usepackage{amssymb}
\usepackage{amsmath}
\usepackage{amsfonts}
\usepackage{mathpartir}
\usepackage[inference]{semantic}
\usepackage{stackengine}
\usepackage{xcolor}
\stackMath
\usepackage{ifthen}
\usepackage{paralist}
\usepackage{hyperref}
\usepackage{xspace}
\usepackage{multirow}
\usepackage{tikz}
\usetikzlibrary{automata, positioning, arrows.meta, arrows, decorations.pathmorphing,shapes,tikzmark}

\pgfarrowsdeclare{biggertip}{biggertip}{  
  \setlength{\arrowsize}{0.4pt}  
  \addtolength{\arrowsize}{.5\pgflinewidth}  
  \pgfarrowsrightextend{0}  
  \pgfarrowsleftextend{-5\arrowsize}  
}{  
  \setlength{\arrowsize}{0.4pt}  
  \addtolength{\arrowsize}{.5\pgflinewidth}  
  \pgfpathmoveto{\pgfpoint{-5\arrowsize}{4\arrowsize}}  
  \pgfpathlineto{\pgfpointorigin}  
  \pgfpathlineto{\pgfpoint{-5\arrowsize}{-4\arrowsize}}  
  \pgfusepathqstroke  
}  
\usepackage[normalem]{ulem}

\usepackage{alltt}

\usepackage{pgfplots}
\pgfplotsset{compat=1.16}
\usepackage{varwidth}
\usepackage{relsize}
\tikzset{
	->, %
	>=Stealth, %
	node distance=3cm, %
	every state/.style={thick, fill=gray!10}, %
	initial text=$ $, %
}

\usepackage[textsize=footnotesize,obeyFinal]{todonotes} %

\usepackage{centernot}
\usepackage{upgreek}
\usepackage{rotating}
\usepackage[linesnumbered,ruled,vlined]{algorithm2e}
\SetKwComment{Comment}{/* }{ */}

\SetCommentSty{mycommfont}\usepackage{mathtools}

\usepackage[appendix=append,bibliography=common]{apxproof}

\newtheoremrep{apxtheorem}{Theorem}
\newtheoremrep{apxlemma}{Lemma}
\newtheoremrep{apxproposition}{Proposition}
\newtheoremrep{apxcorollary}{Corollary}

\usepackage{graphicx}

\newcounter{sqindex}

\newtheoremrep{theorem}{Theorem}
\newtheoremrep{lemma}{Lemma}
\newtheoremrep{proposition}{Proposition}

\makeatletter %
\def\arcr{\@arraycr}
\makeatother

\makeatletter
\newcommand{\xdashrightarrow}[2][]{\ext@arrow 0359\rightarrowfill@@{#1}{#2}}
\newcommand{\xdashleftarrow}[2][]{\ext@arrow 3095\leftarrowfill@@{#1}{#2}}
\newcommand{\xdashleftrightarrow}[2][]{\ext@arrow 3359\leftrightarrowfill@@{#1}{#2}}
\def\rightarrowfill@@{\arrowfill@@\relax\relbar\rightarrow}
\def\leftarrowfill@@{\arrowfill@@\leftarrow\relbar\relax}
\def\leftrightarrowfill@@{\arrowfill@@\leftarrow\relbar\rightarrow}
\def\arrowfill@@#1#2#3#4{%
	$\m@th\thickmuskip0mu\medmuskip\thickmuskip\thinmuskip\thickmuskip
	\relax#4#1
	\xleaders\hbox{$#4#2$}\hfill
	#3$%
}
\makeatother

\renewcommand{\date}{{DATE}\xspace}

\newcommand\ltl[1][]{%
\ifthenelse{\equal{#1}{}}
  {{\textrm{LTL}}}
  {{\textrm{LTL}$(#1)$}}%
\xspace
}

\newcommand\abs[2][]{%
\ifthenelse{\equal{#1}{}}
  {{\alpha({#2})}}
  {{\alpha_{_{#1}}({#2})}}
}

\newcounter{sarrow}

\newcommand{\citet}[1]{\cite{#1}}

\newcommand{\sweap}{\texttt{sweap}\xspace}
\newcommand{\issy}{\texttt{Issy}\xspace}
\newcommand{\issyl}{\texttt{ISSY}\xspace}
\newcommand{\rpg}{\texttt{RPG}\xspace}
\newcommand{\tsl}{\texttt{TSL}\xspace}

\usepackage{listings}
\usepackage{xcolor}

\definecolor{keywordcolor}{rgb}{0.0, 0.0, 0.8}
\definecolor{sectioncolor}{rgb}{0.6, 0.0, 0.6}
\definecolor{typecolor}{rgb}{0.0, 0.6, 0.0}
\definecolor{operatorcolor}{rgb}{0.8, 0.4, 0.0}
\definecolor{commentcolor}{rgb}{0.5, 0.5, 0.5}
\definecolor{stringcolor}{rgb}{0.8, 0.0, 0.0}
\definecolor{backgroundcolor}{rgb}{0.97, 0.97, 0.97}

\lstdefinelanguage{sweap}{
    morekeywords=[1]{INPUTS, OUTPUTS, STATE, VARIABLES, CONTROL, STATES, OBJECTIVE, TRANSITIONS},
    keywordstyle=[1]\color{sectioncolor}\bfseries,
    morekeywords=[2]{integer, natural, boolean, real},
    keywordstyle=[2]\color{typecolor}\bfseries,
    morekeywords=[3]{},
    keywordstyle=[3]\color{keywordcolor}\bfseries,
    otherkeywords={->, ==, >=, <=, !=, :=, &&, ||, !, +, -, *, /, \%, <, >},
    alsoletter={_},
    morestring=[b]",
    morestring=[b]',
    morecomment=[l]{//},
    morecomment=[s]{/*}{*/},
    sensitive=true,
}

\lstdefinestyle{sweapstyle}{
    language=sweap,
    backgroundcolor=\color{backgroundcolor},
    basicstyle=\ttfamily\small,
    keywordstyle=\color{keywordcolor}\bfseries,
    stringstyle=\color{stringcolor},
    commentstyle=\color{commentcolor}\itshape,
    frame=single,
    framesep=5pt,
    framexleftmargin=5pt,
    xleftmargin=10pt,
    numbers=left,
    numberstyle=\tiny\color{gray},
    numbersep=8pt,
    stepnumber=1,
    breaklines=true,
    breakatwhitespace=false,
    tabsize=4,
    showspaces=false,
    showstringspaces=false,
    showtabs=false,
    literate=
        {->}{{{\color{operatorcolor}$\rightarrow$}}}2
        {:=}{{{\color{operatorcolor}:=}}}2
        {>=}{{{\color{operatorcolor}$\geq$}}}2
        {<=}{{{\color{operatorcolor}$\leq$}}}2
        {==}{{{\color{operatorcolor}==}}}2
        {!=}{{{\color{operatorcolor}$\neq$}}}2
        {&&}{{{\color{operatorcolor}\&\&}}}2
        {||}{{{\color{operatorcolor}$\|$$\|$}}}2
}

\lstdefinestyle{sweapminimal}{
    language=sweap,
    basicstyle=\ttfamily\small,
    keywordstyle=[1]\bfseries,
    keywordstyle=[2]\itshape,
    keywordstyle=[3]\underbar,
    frame=lines,
    numberstyle=\tiny,
    breaklines=true,
    tabsize=2,
    escapeinside={(*@}{@*)}
}

\lstdefinestyle{sweapenhanced}{
    style=sweapstyle,
    mathescape=true,
    literate=
        {INPUTS}{{{\color{sectioncolor}\textbf{INPUTS}}}}6
        {OUTPUTS}{{{\color{sectioncolor}\textbf{OUTPUTS}}}}7
        {STATE}{{{\color{sectioncolor}\textbf{STATE}}}}5
        {VARIABLES}{{{\color{sectioncolor}\textbf{VARIABLES}}}}9
        {CONTROL}{{{\color{sectioncolor}\textbf{CONTROL}}}}7
        {STATES}{{{\color{sectioncolor}\textbf{STATES}}}}6
        {OBJECTIVE}{{{\color{sectioncolor}\textbf{OBJECTIVE}}}}9
        {TRANSITIONS}{{{\color{sectioncolor}\textbf{TRANSITIONS}}}}11
        {->}{{{\color{operatorcolor}$\rightarrow$}}}2
        {:=}{{{\color{operatorcolor}:=}}}2
        {>=}{{{\color{operatorcolor}$\geq$}}}2
        {<=}{{{\color{operatorcolor}$\leq$}}}2
        {==}{{{\color{operatorcolor}=}}}2
        {!=}{{{\color{operatorcolor}$\neq$}}}2
        {&&}{{{\color{operatorcolor}$\land$}}}2
        {||}{{{\color{operatorcolor}$\lor$}}}2
}

%% file: tables/table_sweap-pf_sweap-strix.tex
\begin{tabular}[t]{llll}
\toprule
Name & R & Sweap-strix (s) & Sweap (portfolio) (s) \\
\midrule
box & $\bullet$ & 15.66 & 22.18 \\
box-limited & $\bullet$ & 4.06 & 4.65 \\
diagonal & $\bullet$ & 3.82 & 4.94 \\
evasion & $\bullet$ & \phantom{O} & \phantom{O} \\
follow & $\bullet$ & \phantom{O} & 18.06 \\
solitary & $\bullet$ & 5.14 & 4.64 \\
square & $\bullet$ & 81.39 & 41.97 \\
g-real & $\bullet$ & 7.67 & 10.3 \\
g-unreal-1 & \phantom{X} & 88.32 & 3.53 \\
g-unreal-2 & \phantom{X} & ERR & 10.92 \\
g-unreal-3 & \phantom{X} & \phantom{O} & 4.59 \\
heim-double-x & $\bullet$ & 53.05 & 25.97 \\
robot-cat-real-1d & $\bullet$ & 7.22 & 7.72 \\
robot-cat-unreal-1d & \phantom{X} & 5.69 & 6.2 \\
robot-cat-real-2d & $\bullet$ & \phantom{O} & \phantom{O} \\
robot-cat-unreal-2d & \phantom{X} & \phantom{O} & 28.74 \\
robot-grid-reach-1d & $\bullet$ & 3.72 & 4.57 \\
robot-grid-reach-2d & $\bullet$ & 5.19 & 7.3 \\
F-G-contradiction-1 & \phantom{X} & \phantom{O} & 4.62 \\
F-G-contradiction-2 & \phantom{X} & 3.31 & 4.26 \\
f-real & $\bullet$ & 6.44 & 7.96 \\
f-unreal & \phantom{X} & 3.31 & 4.98 \\
ordered-visits & $\bullet$ & 5.21 & 8.76 \\
ordered-visits-choice & $\bullet$ & 5.03 & 8.55 \\
precise-reachability & $\bullet$ & 4.66 & 7.14 \\
robot-to-target & $\bullet$ & ERR & 485.69 \\
robot-to-target-unreal & \phantom{X} & \phantom{O} & \phantom{O} \\
robot-to-target-charging & $\bullet$ & ERR & \phantom{O} \\
robot-to-target-charging-unreal & \phantom{X} & ERR & 34.16 \\
thermostat-F & $\bullet$ & 7.04 & 7.96 \\
thermostat-F-unreal & \phantom{X} & 6.03 & 6.58 \\
unordered-visits-charging & $\bullet$ & \phantom{O} & \phantom{O} \\
unordered-visits & $\bullet$ & 6.67 & 7.37 \\
sort4 & $\bullet$ & \phantom{O} & 131.27 \\
sort5 & $\bullet$ & \phantom{O} & \phantom{O} \\
chain-4 & $\bullet$ & 38.84 & 9.19 \\
chain-5 & $\bullet$ & 441.03 & 12.47 \\
chain-6 & $\bullet$ & \phantom{O} & 17.43 \\
chain-7 & $\bullet$ & \phantom{O} & 54.87 \\
chain-simple-5 & $\bullet$ & 5.01 & 6.66 \\
chain-simple-10 & $\bullet$ & 6.25 & 8.5 \\
chain-simple-20 & $\bullet$ & 12.73 & 11.37 \\
chain-simple-30 & $\bullet$ & 19.77 & 17.97 \\
chain-simple-40 & $\bullet$ & 33.32 & 22.37 \\
chain-simple-50 & $\bullet$ & 54.18 & 35.34 \\
chain-simple-60 & $\bullet$ & 68.06 & 39.64 \\
chain-simple-70 & $\bullet$ & 93.89 & 53.23 \\
\bottomrule
\end{tabular}
\begin{tabular}[t]{llll}
\toprule
Name & R & Sweap-strix (s) & Sweap (portfolio) (s) \\
\midrule
items-processing & $\bullet$ & 17.27 & 12.99 \\
robot-analyze & $\bullet$ & 8.71 & 7.91 \\
robot-collect-v1 & $\bullet$ & 5.61 & 7.19 \\
robot-collect-v2 & $\bullet$ & 7.06 & 7.24 \\
robot-collect-v3 & $\bullet$ & 27.94 & 12.55 \\
robot-deliver-v1 & $\bullet$ & 8.58 & 9.17 \\
robot-deliver-v2 & $\bullet$ & 19.37 & 10.34 \\
robot-deliver-v3 & $\bullet$ & 118.29 & 30.26 \\
robot-deliver-v4 & $\bullet$ & 43.29 & 15.99 \\
robot-deliver-v5 & $\bullet$ & 46.7 & 16.58 \\
robot-repair & \phantom{X} & \phantom{O} & 6.69 \\
robot-running & $\bullet$ & 33.27 & 12.87 \\
scheduler & $\bullet$ & 5.5 & 8.23 \\
heim-buechi & $\bullet$ & 4.86 & 5.69 \\
heim-fig7 & \phantom{X} & 3.15 & 4.11 \\
robot-commute-1d & $\bullet$ & 4.98 & 7.17 \\
robot-commute-2d & $\bullet$ & 55.25 & 135.34 \\
robot-resource-1d & \phantom{X} & 6.8 & 4.3 \\
robot-resource-2d & \phantom{X} & \phantom{O} & 5.69 \\
buffer-storage & $\bullet$ & 4.89 & 4.88 \\
gf-real & $\bullet$ & 3.57 & 4.5 \\
gf-unreal & \phantom{X} & \phantom{O} & 4.19 \\
GF-G-contradiction & \phantom{X} & \phantom{O} & 4.72 \\
helipad & $\bullet$ & 6.29 & 7.57 \\
helipad-contradict & \phantom{X} & \phantom{O} & 6.37 \\
package-delivery & $\bullet$ & 26.79 & 16.09 \\
patrolling & $\bullet$ & 4.23 & 5.39 \\
patrolling-alarm & $\bullet$ & 3.81 & 5.27 \\
storage-GF-64 & $\bullet$ & 3.86 & 5.11 \\
tasks & $\bullet$ & 4.39 & 5.22 \\
tasks-unreal & \phantom{X} & \phantom{O} & 7.36 \\
thermostat-GF & $\bullet$ & 24.5 & 50.38 \\
thermostat-GF-unreal & \phantom{X} & \phantom{O} & 7.74 \\
arbiter & $\bullet$ & 3.57 & 5.43 \\
arbiter-failure & $\bullet$ & 3.39 & 4.23 \\
elevator & $\bullet$ & 3.23 & 5.34 \\
infinite-race & $\bullet$ & 3.37 & 4.22 \\
infinite-race-u & \phantom{X} & \phantom{O} & 4.72 \\
infinite-race-unequal-1 & $\bullet$ & 5.94 & 9.46 \\
infinite-race-unequal-2 & $\bullet$ & ERR & \phantom{O} \\
reversible-lane-r & $\bullet$ & 5.18 & 8.18 \\
reversible-lane-u & \phantom{X} & 11.95 & 5.4 \\
rep-reach-obst-1d & $\bullet$ & 3.61 & 4.58 \\
rep-reach-obst-2d & $\bullet$ & 5.76 & 10.85 \\
rep-reach-obst-6d & $\bullet$ & \phantom{O} & \phantom{O} \\
robot-collect-v4 & $\bullet$ & 16.02 & 9.89 \\
taxi-service & $\bullet$ & 34.45 & 38.09 \\
taxi-service-u & \phantom{X} & 4.89 & 5.42 \\
\bottomrule
\end{tabular}

%% file: tables/table_sweap-issy-pf_issy3.tex
\begin{tabular}[t]{llll}
\toprule
Name & R & Issy (s) & Sweap (s) \\
\midrule
heim-fig7 & \phantom{X} & 0.41 & 6.16 \\
arbiter-with-failure-variant & \phantom{X} & ERR & 4.97 \\
elevator & $\bullet$ & ERR & 6.08 \\
infinite-race & $\bullet$ & \phantom{O} & 6.87 \\
infinite-race-u & \phantom{X} & 24.96 & 4.78 \\
infinite-race-unequal-1-variant & \phantom{X} & 8.0 & 5.45 \\
infinite-race-unequal-2 & $\bullet$ & \phantom{O} & \phantom{O} \\
reversible-lane-r-variant & \phantom{X} & ERR & 5.72 \\
reversible-lane-u & \phantom{X} & \phantom{O} & 5.71 \\
rep-reach-obst-1d & $\bullet$ & \phantom{O} & 5.47 \\
rep-reach-obst-2d & $\bullet$ & \phantom{O} & 11.47 \\
rep-reach-obst-6d & $\bullet$ & \phantom{O} & \phantom{O} \\
taxi-service & $\bullet$ & \phantom{O} & 43.86 \\
taxi-service-u & \phantom{X} & 11.72 & 7.27 \\
balancer-bool-simplified-1 & $\bullet$ & ERR & \phantom{O} \\
balancer-bool-simplified-2 & $\bullet$ & \phantom{O} & \phantom{O} \\
balancer-bool-simplified-3 & $\bullet$ & \phantom{O} & \phantom{O} \\
balancer & $\bullet$ & ERR & \phantom{O} \\
fig7-gt1 & \phantom{X} & 0.61 & 6.43 \\
two-loc-inp-real & $\bullet$ & 0.64 & 6.52 \\
two-loc-inp-unreal-0 & \phantom{X} & 0.48 & 6.61 \\
two-loc-inp-unreal-1 & \phantom{X} & 0.49 & 5.02 \\
two-loc-real-1 & $\bullet$ & 0.34 & 6.05 \\
two-loc-real-2 & $\bullet$ & 0.43 & 6.42 \\
two-vars-real & $\bullet$ & 4.47 & 7.54 \\
two-vars-unreal & \phantom{X} & 0.61 & 6.52 \\
counter-10-10-formula & $\bullet$ & \phantom{O} & OOM \\
counter-10-10-game & $\bullet$ & \phantom{O} & \phantom{O} \\
counter-2-10-2-formula & $\bullet$ & \phantom{O} & 8.35 \\
counter-2-10-2-game & $\bullet$ & \phantom{O} & 69.34 \\
counter-2-10-formula & $\bullet$ & \phantom{O} & 8.58 \\
counter-2-10-game & $\bullet$ & \phantom{O} & 63.72 \\
counter-3-10-formula & $\bullet$ & \phantom{O} & 43.71 \\
counter-3-10-game & $\bullet$ & \phantom{O} & \phantom{O} \\
counter-3-7-formula & $\bullet$ & \phantom{O} & 40.73 \\
counter-3-7-game & $\bullet$ & \phantom{O} & \phantom{O} \\
parity-two-vars-real & $\bullet$ & 5.99 & 6.43 \\
parity-two-vars-unreal-0 & \phantom{X} & 2.23 & 6.3 \\
parity-two-vars-unreal-1 & \phantom{X} & 1.79 & 5.81 \\
parity-two-vars-unreal-2 & \phantom{X} & 0.61 & 5.34 \\
balance-add-rem-2-2-8 & \phantom{X} & 188.51 & \phantom{O} \\
balance-add-rem-8-1-16 & $\bullet$ & 1.83 & \phantom{O} \\
balance-add-rem-8-1-7 & \phantom{X} & 19.71 & \phantom{O} \\
balance-add-rem-8-2-16 & \phantom{X} & 120.25 & OOM \\
empty-add-rem-2-1-unreal & \phantom{X} & ERR & \phantom{O} \\
empty-add-rem-2-1 & $\bullet$ & \phantom{O} & \phantom{O} \\
empty-balance-add-rem-2-1-2 & $\bullet$ & \phantom{O} & \phantom{O} \\
test-01 & \phantom{X} & 2.08 & 9.87 \\
test-02 & $\bullet$ & 21.94 & 14.53 \\
test-03 & $\bullet$ & 479.53 & 18.88 \\
test-04 & $\bullet$ & 98.05 & 18.25 \\
test-05 & $\bullet$ & 5.87 & 43.1 \\
test-06 & \phantom{X} & 0.42 & 4.09 \\
test-07 & $\bullet$ & \phantom{O} & 9.47 \\
test-08 & $\bullet$ & 37.77 & 10.59 \\
test-09 & $\bullet$ & 130.61 & 5.68 \\
test-10 & \phantom{X} & 0.62 & 4.68 \\
\bottomrule
\end{tabular}
\begin{tabular}[t]{llll}
\toprule
Name & R & Issy (s) & Sweap (s) \\
\midrule
test-11 & $\bullet$ & 38.85 & 4.75 \\
test-12 & $\bullet$ & 6.75 & 5.04 \\
test-13 & $\bullet$ & 2.56 & 3.67 \\
test-14 & $\bullet$ & 8.94 & 23.49 \\
test-extract-input & $\bullet$ & 0.28 & 4.32 \\
test-extract-lemma & $\bullet$ & 0.51 & 4.59 \\
buchi-hard-loops-200 & \phantom{X} & 86.13 & 52.41 \\
buchi-hard-loops-400 & \phantom{X} & 312.16 & 220.18 \\
buchi-hard-loops-800 & \phantom{X} & \phantom{O} & 227.66 \\
buchi & $\bullet$ & 33.46 & 32.16 \\
buchi-loops-200 & $\bullet$ & 37.74 & 6.31 \\
buchi-loops-400 & $\bullet$ & 55.8 & 5.61 \\
buchi-loops-800 & $\bullet$ & 125.49 & 5.21 \\
buchi-loops-chaining-200 & $\bullet$ & 542.59 & 19.29 \\
buchi-loops-chaining-400 & $\bullet$ & \phantom{O} & 37.93 \\
buchi-loops-chaining-800 & $\bullet$ & \phantom{O} & 78.36 \\
buchi-loops-swap-200 & $\bullet$ & \phantom{O} & 90.5 \\
buchi-loops-swap-400 & $\bullet$ & \phantom{O} & 25.18 \\
buchi-loops-swap-800 & $\bullet$ & \phantom{O} & 584.69 \\
buchi-simple & $\bullet$ & 13.03 & 20.85 \\
choice-3-actions & $\bullet$ & 145.62 & \phantom{O} \\
choice-4-actions & $\bullet$ & \phantom{O} & 22.62 \\
choice-actions-i & $\bullet$ & 480.89 & 14.4 \\
double & $\bullet$ & \phantom{O} & \phantom{O} \\
equality-assumption & $\bullet$ & \phantom{O} & 16.05 \\
equal-mod2-real & $\bullet$ & 1.83 & 5.15 \\
equal-mod2-unreal & \phantom{X} & \phantom{O} & \phantom{O} \\
fault-tolerance & $\bullet$ & \phantom{O} & 5.99 \\
inequality-assumption & $\bullet$ & 14.6 & 7.52 \\
lemma-chaining-control & $\bullet$ & \phantom{O} & 7.43 \\
lemma-chaining & $\bullet$ & 52.88 & 5.37 \\
nested-x-y-z & $\bullet$ & \phantom{O} & 6.22 \\
nested-x-y-z-u & $\bullet$ & \phantom{O} & 10.85 \\
nested-x-y-z-u-v & $\bullet$ & \phantom{O} & 56.51 \\
nondet-exit-swap-input & $\bullet$ & 350.58 & 7.82 \\
nondet-exit-swap & $\bullet$ & 2.31 & 4.9 \\
prioritized-tasks-real-100 & $\bullet$ & \phantom{O} & 10.73 \\
prioritized-tasks-real-200 & $\bullet$ & \phantom{O} & 10.08 \\
prioritized-tasks-unreal-100 & \phantom{X} & 17.12 & 459.84 \\
ranking-choice-2 & $\bullet$ & 21.27 & 6.84 \\
ranking-choice-3 & $\bullet$ & 109.7 & 20.31 \\
ranking-choice-4 & $\bullet$ & \phantom{O} & \phantom{O} \\
reach-either-or & $\bullet$ & 109.11 & 5.77 \\
service-10 & $\bullet$ & \phantom{O} & 7.31 \\
service-2 & $\bullet$ & 22.67 & 6.34 \\
service-4 & $\bullet$ & 288.59 & 6.09 \\
service-8 & $\bullet$ & \phantom{O} & 8.97 \\
service-bounds-10 & $\bullet$ & 259.24 & 5.12 \\
service-bounds-20 & $\bullet$ & 257.8 & 5.86 \\
ex-2-05 & $\bullet$ & 1.03 & 3.92 \\
ex-4-06 & $\bullet$ & 2.04 & 5.41 \\
ex-4-11 & $\bullet$ & \phantom{O} & 5.1 \\
ex-4-12 & \phantom{X} & \phantom{O} & \phantom{O} \\
ex-4-13 & $\bullet$ & \phantom{O} & \phantom{O} \\
ex-4-15 & $\bullet$ & \phantom{O} & \phantom{O} \\
ex-4-18 & $\bullet$ & 259.08 & 5.46 \\
ex-4-21 & $\bullet$ & 1.17 & 3.87 \\
ex-5-01 & $\bullet$ & \phantom{O} & 7.64 \\
\bottomrule
\end{tabular}

%% file: tables/table_sweap-rpg-pf_issy3-rpg.tex
\begin{tabular}[t]{llll}
\toprule
Name & R & Issy (RPG) (s) & Sweap (s) \\
\midrule
box & $\bullet$ & 0.48 & 12.36 \\
box-limited & $\bullet$ & 0.54 & 4.62 \\
diagonal & $\bullet$ & 0.43 & 9.94 \\
evasion & $\bullet$ & 0.71 & 11.63 \\
follow & $\bullet$ & \phantom{O} & 12.23 \\
solitary & $\bullet$ & 0.28 & 4.67 \\
square & $\bullet$ & 0.55 & 39.86 \\
g-real & $\bullet$ & 533.75 & 45.29 \\
g-unreal-1 & \phantom{X} & 0.3 & 5.71 \\
g-unreal-2 & \phantom{X} & 0.69 & 11.41 \\
g-unreal-3 & \phantom{X} & 0.38 & 4.81 \\
heim-normal & $\bullet$ & 0.73 & 4.7 \\
heim-double-x & $\bullet$ & 7.53 & \phantom{O} \\
robot-cat-real-1d & $\bullet$ & \phantom{O} & \phantom{O} \\
robot-cat-unreal-1d & \phantom{X} & \phantom{O} & 5.91 \\
robot-cat-real-2d & $\bullet$ & \phantom{O} & \phantom{O} \\
robot-cat-unreal-2d & \phantom{X} & \phantom{O} & 11.62 \\
robot-grid-reach-1d & $\bullet$ & 0.43 & 4.81 \\
robot-grid-reach-2d & $\bullet$ & 1.07 & 7.02 \\
F-G-contradiction-1 & \phantom{X} & 0.24 & 4.17 \\
F-G-contradiction-2 & \phantom{X} & 2.27 & 7.36 \\
f-real & $\bullet$ & 0.83 & 5.74 \\
f-unreal & \phantom{X} & 2.06 & 8.85 \\
ordered-visits & $\bullet$ & 25.69 & 18.98 \\
ordered-visits-choice & $\bullet$ & 10.63 & 15.11 \\
precise-reachability & $\bullet$ & 5.7 & 5.66 \\
robot-to-target & $\bullet$ & \phantom{O} & 243.51 \\
robot-to-target-unreal & \phantom{X} & 138.25 & 71.21 \\
robot-to-target-charging & $\bullet$ & \phantom{O} & 115.54 \\
robot-to-target-charging-unreal & \phantom{X} & \phantom{O} & 113.7 \\
thermostat-F & $\bullet$ & 13.82 & 8.68 \\
thermostat-F-unreal & \phantom{X} & 5.48 & 6.1 \\
unordered-visits-charging & $\bullet$ & \phantom{O} & \phantom{O} \\
unordered-visits & $\bullet$ & 17.31 & 14.05 \\
sort4 & $\bullet$ & 61.17 & \phantom{O} \\
sort5 & $\bullet$ & \phantom{O} & \phantom{O} \\
robot-tasks & $\bullet$ & 10.8 & 4.23 \\
chain-4 & $\bullet$ & 3.01 & 8.55 \\
chain-5 & $\bullet$ & 4.85 & 11.04 \\
chain-6 & $\bullet$ & 5.5 & 17.6 \\
chain-7 & $\bullet$ & 7.88 & 50.04 \\
chain-simple-5 & $\bullet$ & 4.12 & 6.08 \\
chain-simple-10 & $\bullet$ & 15.15 & 8.54 \\
\bottomrule
\end{tabular}
\begin{tabular}[t]{llll}
\toprule
Name & R & Issy (RPG) (s) & Sweap (s) \\
\midrule
chain-simple-20 & $\bullet$ & 51.1 & 9.45 \\
chain-simple-30 & $\bullet$ & 103.58 & 13.53 \\
chain-simple-40 & $\bullet$ & 163.36 & 17.5 \\
chain-simple-50 & $\bullet$ & 264.57 & 25.61 \\
chain-simple-60 & $\bullet$ & 424.87 & 37.85 \\
chain-simple-70 & $\bullet$ & 550.7 & 46.25 \\
items-processing & $\bullet$ & \phantom{O} & 10.09 \\
robot-analyze & $\bullet$ & 3.18 & 7.95 \\
robot-collect-v1 & $\bullet$ & 1.53 & 6.69 \\
robot-collect-v2 & $\bullet$ & 2.38 & 6.6 \\
robot-collect-v3 & $\bullet$ & 1.53 & 6.31 \\
robot-deliver-v1 & $\bullet$ & 3.37 & 8.06 \\
robot-deliver-v2 & $\bullet$ & 4.13 & 16.65 \\
robot-deliver-v3 & $\bullet$ & 3.75 & 132.74 \\
robot-deliver-v4 & $\bullet$ & 4.08 & 50.2 \\
robot-deliver-v5 & $\bullet$ & 3.85 & 68.52 \\
robot-repair & \phantom{X} & 0.46 & 7.42 \\
robot-running & $\bullet$ & 1.77 & 7.55 \\
scheduler & $\bullet$ & 9.4 & 8.46 \\
heim-buechi & $\bullet$ & 1.48 & 6.2 \\
heim-fig7 & \phantom{X} & 0.49 & 5.17 \\
robot-commute-1d & $\bullet$ & 1.11 & 6.32 \\
robot-commute-2d & $\bullet$ & 3.17 & 135.82 \\
robot-resource-1d & \phantom{X} & 0.92 & 4.91 \\
robot-resource-2d & \phantom{X} & 2.61 & 5.88 \\
buffer-storage & $\bullet$ & \phantom{O} & 42.56 \\
gf-real & $\bullet$ & 0.46 & 5.15 \\
gf-unreal & \phantom{X} & 0.31 & 4.92 \\
GF-G-contradiction & \phantom{X} & 0.33 & 5.1 \\
helipad & $\bullet$ & \phantom{O} & 22.97 \\
helipad-contradict & \phantom{X} & 0.45 & 28.15 \\
package-delivery & $\bullet$ & 102.67 & 134.01 \\
patrolling & $\bullet$ & 18.55 & 7.63 \\
patrolling-alarm & $\bullet$ & 23.18 & 12.68 \\
storage-GF-64 & $\bullet$ & 38.36 & 10.35 \\
tasks & $\bullet$ & \phantom{O} & 32.5 \\
tasks-unreal & \phantom{X} & \phantom{O} & 21.42 \\
thermostat-GF & $\bullet$ & 39.43 & 12.15 \\
thermostat-GF-unreal & \phantom{X} & 11.61 & 10.04 \\
arbiter & $\bullet$ & \phantom{O} & \phantom{O} \\
arbiter-failure & $\bullet$ & 5.56 & 436.0 \\
elevator & $\bullet$ & \phantom{O} & \phantom{O} \\
taxi-service & $\bullet$ & OOM & \phantom{O} \\
\bottomrule
\end{tabular}

%% file: tables/table_sweap-tsl-pf_issy3-tsl.tex
\begin{tabular}{llll}
\toprule
Name & R & Issy (TSL) (s) & Sweap (s) \\
\midrule
box & $\bullet$ & 1.08 & 6.78 \\
box-limited & $\bullet$ & 0.78 & 4.69 \\
diagonal & $\bullet$ & 0.93 & 6.12 \\
evasion & $\bullet$ & 2.02 & 8.79 \\
follow & $\bullet$ & 133.93 & 12.19 \\
solitary & $\bullet$ & 0.41 & 4.69 \\
square & $\bullet$ & 2.53 & 12.16 \\
g-real & $\bullet$ & \phantom{O} & 9.07 \\
g-unreal-1 & \phantom{X} & 126.62 & 4.36 \\
g-unreal-2 & \phantom{X} & 0.76 & 11.89 \\
g-unreal-3 & \phantom{X} & 0.39 & 4.76 \\
heim-normal & $\bullet$ & 0.95 & 5.36 \\
heim-double-x & $\bullet$ & 22.62 & \phantom{O} \\
robot-cat-real-1d & $\bullet$ & \phantom{O} & 16.67 \\
robot-cat-unreal-1d & \phantom{X} & \phantom{O} & \phantom{O} \\
robot-cat-real-2d & $\bullet$ & \phantom{O} & 6.06 \\
robot-cat-unreal-2d & \phantom{X} & \phantom{O} & \phantom{O} \\
robot-grid-reach-1d & $\bullet$ & 134.57 & 4.89 \\
robot-grid-reach-2d & $\bullet$ & 2.12 & 9.81 \\
F-G-contradiction-1 & \phantom{X} & 0.3 & 4.64 \\
F-G-contradiction-2 & \phantom{X} & 2.17 & 4.26 \\
f-real & $\bullet$ & 0.74 & 4.23 \\
f-unreal & \phantom{X} & 2.17 & 4.87 \\
ordered-visits & $\bullet$ & 24.78 & 8.22 \\
ordered-visits-choice & $\bullet$ & 11.69 & 9.7 \\
precise-reachability & $\bullet$ & 4.87 & 5.64 \\
robot-to-target & $\bullet$ & 30.92 & 23.03 \\
robot-to-target-unreal & \phantom{X} & 6.3 & 12.61 \\
robot-to-target-charging & $\bullet$ & \phantom{O} & \phantom{O} \\
robot-to-target-charging-unreal & \phantom{X} & \phantom{O} & 15.67 \\
thermostat-F & $\bullet$ & 11.7 & 6.72 \\
thermostat-F-unreal & \phantom{X} & 10.21 & 4.78 \\
unordered-visits-charging & $\bullet$ & \phantom{O} & OOM \\
unordered-visits & $\bullet$ & 135.94 & 8.53 \\
sort4 & $\bullet$ & ERR & \phantom{O} \\
sort5 & $\bullet$ & ERR & \phantom{O} \\
robot-tasks & $\bullet$ & 24.94 & 8.12 \\
heim-buechi & $\bullet$ & 14.55 & 15.76 \\
heim-fig7 & \phantom{X} & 0.78 & 5.01 \\
robot-commute-1d & $\bullet$ & 1.73 & 14.55 \\
robot-commute-2d & $\bullet$ & 10.56 & \phantom{O} \\
robot-resource-1d & \phantom{X} & 1.27 & 6.12 \\
robot-resource-2d-real & $\bullet$ & 3.01 & 21.66 \\
buffer-storage & $\bullet$ & \phantom{O} & 9.3 \\
gf-real & $\bullet$ & 0.47 & 4.96 \\
gf-unreal & \phantom{X} & 0.36 & 4.01 \\
GF-G-contradiction & \phantom{X} & 0.39 & 4.55 \\
helipad & $\bullet$ & \phantom{O} & 6.55 \\
helipad-contradict & \phantom{X} & 53.21 & 6.0 \\
package-delivery & $\bullet$ & 209.84 & 8.07 \\
patrolling & $\bullet$ & 18.26 & 5.91 \\
patrolling-alarm & $\bullet$ & 22.96 & 5.66 \\
storage-GF-64 & $\bullet$ & 55.63 & 6.09 \\
tasks & $\bullet$ & 42.58 & 10.73 \\
tasks-unreal & \phantom{X} & \phantom{O} & 9.18 \\
thermostat-GF & $\bullet$ & 32.74 & 7.46 \\
thermostat-GF-unreal & \phantom{X} & 14.18 & 5.8 \\
arbiter & $\bullet$ & \phantom{O} & 6.59 \\
arbiter-failure & $\bullet$ & 1.6 & 5.76 \\
elevator & $\bullet$ & ERR & 13.43 \\
reversible-lane-r-variant & \phantom{X} & OOM & OOM \\
reversible-lane-u & \phantom{X} & \phantom{O} & 93.76 \\
rep-reach-obst-1d & $\bullet$ & ERR & 6.03 \\
rep-reach-obst-2d & $\bullet$ & 121.87 & 20.59 \\
taxi-service & $\bullet$ & ERR & \phantom{O} \\
\bottomrule
\end{tabular}